\documentclass[reprint,amsmath,amssymb,graphicx,longbibliography,showpacs]{revtex4-1}

\usepackage{bm}
\usepackage{stmaryrd}
\usepackage{amsmath}
\usepackage{calligra}
\usepackage{mathtools}
\usepackage{amsfonts}
\usepackage{amssymb}
\usepackage{textcomp}
\usepackage{graphicx}
\usepackage{caption}
\usepackage[margin=15mm]{geometry}
\usepackage{yfonts}
\usepackage[breaklinks=true,colorlinks=true,linkcolor=blue,urlcolor=blue,citecolor=blue]{hyperref}
\usepackage[lofdepth,lotdepth]{subfig}

\begin{document}

\title{Entanglement of two hybrid optomechanical cavities composed of BEC atoms under Bell detection }
\author{M. Eghbali-Arani$^{1}$} 
\author{V. Ameri$^{2}$}
\affiliation{
$^1$Department of Physics, University of Kashan, Kashan, Iran \\
$^2$Department of Physics, Faculty of Science, University of Hormozgan, Bandar-Abbas, Iran}
\begin{abstract}
In this paper, firstly, we consider bipartite entanglement between each part of an optomechanical cavity composed of one dimensional Bose-Einstein condensate (BEC). We investigate atomic collision on the behavior of the BEC in the week photon-atom coupling constant, and use Bogoliubov approximation for the BEC. Secondly under above condition, we propose a scheme for entanglement swapping protocol wich involves tripartite systems. In our investigation, we consider a scenario where BECs, moving mirrors, and optical cavity modes are given in a Gaussian state with a covariance matrix (CM). By applying the Bell measurement to the output optical field modes, we show how the remote entanglement between two BECs, two moving mirrors, and BEC-mirror modes in different optomechanical cavity can be generated.
\end{abstract}

\maketitle

\section{Introduction}
In recent years, non-classical entanglement states play an essential role for the communication and computation processing \cite{RevModPhys.77.513}. Entangled states of continuous variable (CV) systems are attractive for this purpose \cite{RevModPhys.81.865,weedbrook2012gaussian}. As an example of continuous variable systems, Gaussian states play a key role in the quantum information since thay have been formulated easily and they can be created and controlled exprimentally.\\
Due to the extensive application of quantum entanglement, several systems have been proposed for generating entanglement between two nodes of a quantum system \cite{PhysRevLett.82.1975, PhysRevLett.81.3631, PhysRevLett.85.2392,Mendonça201479, Banerjee2010816}. Entanglement swapping is one of the non-locality effects for generating quantum correlation between two non-interacting distant systems \cite{PhysRevLett.80.3891,deLimaBernardo201680,Haq2015290}. This technique experimentally and theoretically has been studied in Refs \cite{PhysRevA.89.022331, PhysRevLett.109.143601, PhysRevLett.93.250503, PhysRevLett.94.220502}. \\  
For nontrivial quantum communication tasks such as teleportation, entanglement is necessary to teleport the information between two remote channels. So far, scientists have studied some schemes for the generating entanglement between two distant systems \cite{PhysRevA.89.022331, PhysRevLett.109.143601, PhysRevLett.93.250503, PhysRevLett.94.220502}. Therefore, in this work, we propose a hybrid scheme to show entanglement between two distant nodes that never interact. We show a scenario for entangling two remote and initially uncorrelated modes by applying the Bell detection to the output optical field modes. i. e. we present a general scheme to entangled two part of remote systems (two BECs, two mirrors, and BEC-mirror) by using a balanced beam splitter and two homodyne detectors. In simple words, we use the quantum correlations between two optical fields subparts that fully disentangled in order to create entanglement between other part of systems.
 In this scheme, the excitation of the ultracold atoms plays the role of the vibrational mode of the mirror in an optomechanical system \cite{PhysRevA.87.013417}.

This paper is organized as follows. In Sec. \ref{formulation} we provide a brief theoretical description of the system under consideration and then we study the quantum state transferring of output mode and BEC modes inside the one cavity and quantify the entanglement between the output optical modes by using the logarithmic negativity, while in Sec. \ref{entanglement}, we firstly describe the Bell measurement protocol and then we perform an analysis of the entanglement of two remote systems. Finally, our conclusions are given in Sec. \ref{conclusion}

\section{Formulation and Theoretical Description of The System at each node}\label{formulation}
\begin{figure}[ht]
\centering
\includegraphics[width=3in]{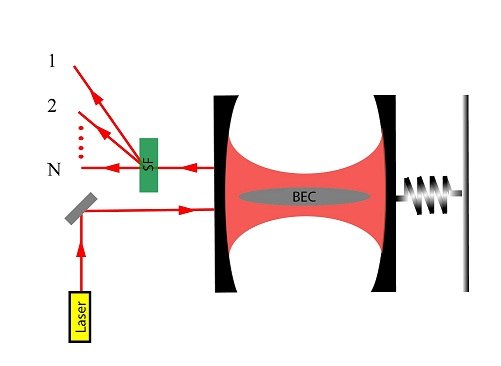}
\caption{(Color online) Trapped BEC atoms inside an optomechanical cavity.}
\label{f1}
\end{figure}

We study a cigar-shaped gas of $N$ ultracold bosonic two-level atoms with transition frequency $\omega_a$ and mass $m$ in a Fabry-Perot cavity with a movable end mirror with the frequency of $\omega_m$ as sketch in Fig. \ref{f1}.  The optical cavity with length $L$, is driving at rate $E$, and the wave number $K=\omega_{l}/c$, where $\omega_l$ is the frequency of laser pump and $c$ is the speed of light.
By assuming that the laser pump $\omega_l$ is far detuned from the atomic transition frequency $\omega_a$, and ignoring spontaneous emission, we can adiabatically eliminated the excited electronic state of the atoms. So
the many-body Hamiltonian of system in the frame rotating at pump frequency is given by \cite{Maschler2004145}

\begin{eqnarray}\label{Hamiltonian}
\hat H&=& \hbar\Delta_{c} {\hat a}^\dagger \hat a+i \hbar E ({\hat a}^\dagger - \hat a)+\frac{1}{2}\hbar\omega_m(\hat p^2+\hat q^2)-\hbar g \hat a^\dagger \hat a \hat q\nonumber\\&+& \int dx {\psi}^\dagger (x) \hat H_{0}\psi(x) + {\hat H_{aa}},
\end{eqnarray}

where $\hat p$ and $\hat q$ are the canonical position and momentum of the moving mirror, $\Delta_{c}=\omega_{c}-\omega_{l}$ is the detuning between laser pump ($\omega_{l}$), and cavity field ($\omega_{c}$) frequences. $\hat a$ is the destruction operator for the cavity photons, and $g=(\omega_c/L)\sqrt{\hbar/M\omega_m}$ is the strength of coupling constant of radiation pressure with mechanical mode, where $M$ is effective mass of the moving mirror. $\hat H_{aa}$ and $\hat H_{0}$ are the interaction Hamiltonian of two atoms, and the single-particle Hamiltonian of an atom inside the periodic lattice generated by standing optical modes respectively

\begin{eqnarray}
\hat H_{0}&=&{\hat p}^2/2M+\hbar U_{0} {\cos^2(Kx)}  {\hat a}^\dagger \hat a,\nonumber\\
\hat H_{aa}&=&\frac{1}{2} U_{s} \int_{-L/2}^{L/2} dx \psi^\dagger(x)\psi^\dagger(x)\psi(x)\psi(x)
\end{eqnarray}
The parameters $U_{0}=g_{0}^2/\Delta_{a}$ and $g_{0}$ are the optical lattice barrier height per photon and the vacuum Rabi frequency respectively. $U_{s}=\frac{4\pi\hbar^2a_{s}}{M}$, where $a_{s}$ is the two-body s-wave scattering length\cite{PhysRevA.72.053417,PhysRevLett.95.260401}.\\
In the weakly interacting regime, under the Bogoliubov approximation the atomic field operator can be expanded as \cite{nagy2009nonlinear}

\begin{eqnarray}\label{psi}
\psi(x)=\sqrt{\frac{N}{L}}+\sqrt{\frac{2}{L}}\cos(2Kx)\hat c,
\end{eqnarray}
where $\hat c$ is Bogoliubov annihilation operator ($[\hat c,\hat c^\dagger]=1$).
By substituting the atomic field operator, Eq.(\ref{psi}), into the Hamiltonian of Eq.(\ref{Hamiltonian}) and introducing the Bogoliubov mode quadratures $\hat Q=(\hat c+\hat c^\dagger)/\sqrt{2}$, and $\hat P=(\hat c-\hat c^\dagger)/i\sqrt{2}$ one obtains the following Hamiltonian \cite{dalafijpb}

\begin{eqnarray}\label{Hamiltonian2}
\hat H&=&\delta_{c}{\hat a}^\dagger \hat a+i\hbar E({\hat a}^\dagger-\hat a)+\frac{1}{2}\hbar\omega_{m}({\hat p}^2+{\hat q}^2)-\hbar g {\hat a}^\dagger \hat a \hat q\nonumber\\&+&\frac{1}{2}\hbar\Omega_{c}({\hat P}^2+{\hat Q}^2)+\hbar G {\hat a}^\dagger \hat a \hat Q+\frac{1}{2}\hbar\omega_{sw}{\hat Q}^2
\end{eqnarray}

Where $ \delta_{c}=\Delta_{c}+\frac{N U_{0}}{2}$ and $\omega_R=\frac{\hbar{K}^2}{2M}$ are the effective Stark shift detuning and the recoil frequency respectively. $\omega_{sw}=\frac{8 \pi \hbar N}{M L {\nu}^2}$  is the scattering frequency and $\nu$ is the waist of the optical potential. $\Omega_{c}=4\omega_{R}+1/2 \omega_{sw}$ is the effective detuning of BEC. $G=\frac{\omega_c}{L}\sqrt{\frac{\hbar}{4\omega_R m_s}}$ is the coupling constant of radiation pressure with Bogoliubov mode, and $m_s=\frac{\hbar{\omega_c}^2}{L^2N{U_0}^2 \omega_R}$ is effective mass of BEC mode.

\subsection{Dynamics}
The quantum stochastic Langevin equations~(QLEs) for the moving mirror, BEC and the cavity field
variables are obtained by adopting the dissipation-fluctuation theorem \cite{drummond1980generalised}
\begin{eqnarray}\label{qlemain}
\hat{ \dot{q}}&=&\omega_m \hat p,\nonumber\\
\hat{ \dot{p}}&=&-\omega_m \hat q+g \hat a^\dagger \hat a-\gamma_m \hat p+\sqrt{2\gamma_m}\hat p_{in},\nonumber\\
\hat{ \dot{P}}&=&-(\Omega_c+\omega_{sw})\hat Q -\gamma_c\hat P-G \hat a^\dagger \hat a+\sqrt{2\gamma_c}\hat Q_{in},\nonumber\\
\hat{ \dot{Q}}&=&\Omega_c \hat P-\gamma_c \hat Q+\sqrt{2\gamma_c}\hat P_{in},\\
\hat{ \dot{a}}&=&-(\mathrm i \delta_c+\kappa)\hat a -\mathrm i (g \hat q -G\hat Q)\hat a+E_{\mathrm{d}}+\sqrt{2\kappa}\hat a_{in}\nonumber,
\end{eqnarray}
where $\gamma_{m}$ is the moving mirror damping rate, $ \gamma_c $ is the dissipation of the collective
density excitations of the BEC inside of optical lattice, and $\hat a_{in}(t)$ is the cavity field input noise which obeys the white-noise correlation functions \cite{drummond1980generalised}
\begin{align}
\langle \hat a_{in}(t) \hat a_{in}^{\dagger}(t^{\prime})\rangle &  =\delta(t-t^{\prime
}),\langle \hat a_{in}^{\dagger}(t)\hat a_{in}(t^{\prime})\rangle=0,\label{coropt}\\
\langle \hat b_{in}(t) \hat b_{in}^{\dagger}(t^{\prime})\rangle &  =\delta(t-t^{\prime
}),\langle\hat b_{in}^{\dagger}(t) \hat b_{in}(t^{\prime})\rangle=0,\nonumber
\end{align}
where we have set $N=[\mathrm{exp}(\hbar\omega_{c}%
/k_{B}T)-1]^{-1}\approx0$, since $\hbar\omega_{c}/k_{B}T\gg1$ at optical frequency. However, $\xi(t)$ is the Brownian noise acting on the moving mirror, with correlation function \cite{pinard1999effective,law1995interaction}
\begin{equation}
\langle \hat \xi(t)\hat \xi(t^{\prime})\rangle=\frac{\gamma_{m}}{\omega_{m}}\int\frac{d\omega}{2\pi}e^{-i\omega(t-t^{\prime})}\omega\Big[\mathrm{coth}%
\big(\frac{\hbar\omega}{2k_{B}T}\big)+1\Big], \label{nois1}
\end{equation}
where $k_{B}$ is the Boltzmann constant, and $T$ the temperature of the reservoir. In a very high mechanical quality factor regime ( $Q=\omega_{m}/\gamma_{m}\rightarrow\infty$), the mechanical noise of mirror is characterized by white thermal noise \cite{benguria1981quantum}
\[
\frac{\langle\hat \xi(t)\hat \xi(t^{\prime})+\hat \xi(t^{\prime})\hat \xi(t)\rangle}{2}\simeq\gamma
_{m}(2\bar{n}_m+1)\delta(t-t^{\prime})~,
\]
with mean excitation number $\bar{n}_m=[\mathrm{exp}(\hbar\omega_{m}/k_{B}T)-1]^{-1}$. $ \hat Q_{in} $ and $\hat  P_{in} $ are the thermal noise inputs for the Bogoliubov mode of BEC which satisfy the 
Markovian correlation functions \cite{dalafi2013nonlinear,zhang2010hamiltonian}
\[
\langle \hat P_{in}(t)\hat P_{in}(t')\rangle=\langle \hat Q_{in}(t) \hat Q_{in}(t')\rangle=2\gamma_c(n_c+1/2)\delta(t-t')
\]
where $n_c=[\mathrm{exp}(\hbar\omega_{B}/k_{B}T_c)-1]^{-1}$ is the number of thermal excitations for the Bogoliubov mode which oscillates with frequency $ \omega_{B}=\sqrt{\Omega_c(\Omega_c+\omega_{sw})} $, and $T_c$ is the effective tempereature of BEC.\\
Quadrature operators of system can be represent as the $\mathbf{u}=[\hat q,\hat p,\hat Q,\hat P, \hat X,\hat Y]^T$,  where the quadratures of the optical cavity are defined as $ \hat X=(\hat a+\hat a^{\dagger})/\sqrt{2} $ and $\hat Y=i(\hat a^{\dagger}-\hat a)/\sqrt{2} $. Since the cavity is pumped by an intense laser then we linearized the QLEs
given in Eq.(\ref{qlemain}) around the mean values, i.e,. $u_j=u_{j,s}+\delta u_j(t)$, where $ u_{j,s} $ are the semiclassical mean values and $\delta u_j(t)$ are fluctuation operators with zero-mean value. The steady state are obtained 
\begin{eqnarray}
q_{s}&=&\frac{g}{\omega_m}{\alpha}^2,\,\,\,p_{s}=0,\nonumber\\
Q_{s}&=&-\frac{G\alpha^2}{\Omega_c+\omega_{sw}+\frac{\gamma_{c}^2}{\Omega_c}}, \\ 
P_{s}&=&\frac{\gamma_{c}}{\Omega_c}Q_s,\nonumber\\
\alpha &=&\frac{E}{\sqrt{\Delta^2+\kappa^2}},\nonumber
\end{eqnarray}
where $\Delta=\delta_{c}-g q_{s}+G Q_{s}$ is the effective detuning and $ \alpha=\alpha^*$ is the mean value of the optical filed mode.
The dynamics of the fluctuation operators , $\delta \textbf{u}(t)= [\delta \hat q, \delta \hat p,\delta \hat Q,\delta \hat P,\delta \hat X,\delta \hat Y]^T $,  are given by the linearized QLEs which one can write as  
\begin{equation}
\delta \dot{\textbf{u}}(t)=\textbf{A} \delta \textbf{u}(t)+\textbf{n}(t) 
\end{equation}
where
\begin{equation}
\textbf{A}=\left(\begin{array}{cccccc}
    0 & \omega_{m} &0 &0& 0&0 \\
  -\omega_{m} & -\gamma_{m} &0&0 & \sqrt{2}g\alpha_{s} &0 \\
   0 & 0 & -\gamma_{c} & \Omega_c &0 &0 \\
   0 & 0 &-(\Omega_c+\omega_{sw})  & -\gamma_{c} & -\sqrt{2}G\alpha_{s}&0\\
   0& 0 &  0 &0& -\kappa &\Delta\\
   \sqrt{2}g\alpha_{s}&0 & -\sqrt{2}G\alpha_{s}& 0&-\Delta &-\kappa
\end{array}\right),
\label{drift}
\end{equation}
is the drift matrix and $\textbf{n}(t)=[0,\gamma_{m}(2n_m+1),\gamma_{c}(2n_B+1),\gamma_{c}(2n_B+1),\kappa, \kappa]^T$
defines the vector of the noises. We note that the current system can reach a steady state after a transient time when all the
eigenvalues of the drift matrix $ \textbf{A} $ have negative real
value according to the \textit{Routh-Hurwitz} criterion \cite{gradshteyn1980tables}.  

\subsection{Optomechanical entanglement of output modes}
As we mentioned in Sec. II, the basic ingredient for creating a quantum link between the two distant system is that the state at each remote node must possess a nonzero entanglement between the two remote mode and a travelling optical mode. i.e by using the travelling output modes, any quantum communications can be implemented rather than intracavity ones. So it is important to study how the entanglement generated within the cavity between the BEC modes and light or mirror and light are transferred to the output field.
Due to the linearized dynamics of the fluctuations and since all noises are Gaussian the steady state is a zero-mean Gaussian state which is fully characterized by the stationary correlation matrix \textbf{V}.
For analysis of entangling of the vibrating mirror and BEC modes with detectable output field of an optical cavity, we use an expression for output optical fields. The output mode of the optical cavity is given by the standard input-output relation $ \hat a_{\mathrm{out}}=\sqrt{2\kappa}\hat a-\hat a_{\mathrm{in}} $. One can also define the selected output
mode by means of the causal filter function $\hat  a^{\mathrm{filt}}=\int_{t_0}^{t}F(t-s)\hat a_{\mathrm{out}}(s) ds$, where the causal filter function
$ F(t) =\sqrt{2/\tau}\mathrm{exp}[-(1/\tau+i\Omega)t]\Theta (t)$ is characterized by central frequency $ \Omega $, bandwidth $ 1/\tau $, and  the Heaviside step function $ \Theta (t) $ \cite{genes2008robust, barzanjeh2013continuous,eghbali2015generating}. 
 In the frequency domain, the stationary CM for the quantum fluctuations of the mirror, BEC,  and the output mode of the optical cavity variables, $\textbf{u}^{\mathrm{filt}}(t)= [\hat q,  \hat p, \hat Q, \hat P, \hat X^{\mathrm{filt}}, \hat Y^{\mathrm{filt}}]^T $, takes the form
\begin{eqnarray}
\mathbf{V}&=&\underset{t\rightarrow\infty}{\lim}\frac{1}{2}\left\langle
u_{i}^{\mathrm{filt}}(t)u_{j}^{\mathrm{filt}}(t)+u_{j}^{\mathrm{filt}}(t)u_{i}^{\mathrm{filt}}(t)\right\rangle\nonumber\\
&=& \int  d\omega\boldsymbol{\Upsilon}(\omega
)\Big(\mathbf{\tilde{M}}(\omega)+\mathbf{P}%
\Big)\nonumber\\
&&\times\mathbf{D}\Big(\mathbf{\tilde{M}}(\omega)^{\dagger
}+\mathbf{P}\Big)\boldsymbol{\Upsilon}^{\dagger}(\omega),
\label{vmat}
\end{eqnarray}
for each node, where
$\tilde{\textbf{M}}(\omega)=(i\omega \textbf{I}+\textbf{A})^{-1}$,  ${\textbf{P}}=\mathrm{Diag}[0,0,0,0,\frac{1}{2\kappa},\frac{1}{2\kappa}]$ and $\boldsymbol{\Upsilon}(\omega
)$ is the Fourier transform of
\begin{equation}
\boldsymbol{\Upsilon}(t)=\left(\begin{array}{cccccc}
    \delta(t) &0&0 &0& 0&0 \\
   0 & \delta(t) &0&0 & 0&0 \\
   0 & 0 & \delta(t) &0 &0 &0\\
   0 & 0 &0&\delta(t)&0&0\\
   0& 0 &  0 &0& \mathcal{R}&-\mathcal{I}\\
   0 &0 & 0& 0&-\mathcal{I}& \mathcal{R}
  \end{array}\right),
\label{T}
\end{equation}
where $\mathcal{R}=\sqrt{2\kappa}\mathrm{Re}[F(t)]$ and $\mathcal{I}=\sqrt{2\kappa}\mathrm{Im}[F(t)]$ are determined by the causal filter functions $F(t)$ with bandwidth $1/\tau$ and central frequency $\Omega$.\\
The bipartite entanglement between the different parts of subsystems is characterized by the logarithmic negativity
\begin{eqnarray}\label{EN}
{E_{\mathcal{N}}}=\max[0,-\ln(2\eta_{-})],
\end{eqnarray}
where $\eta_{-}=2^{-1/2}\left[ {\sigma-\sqrt{\sigma^2-4 \mathrm{det}\textbf{V}^\prime}}\right] ^{1/2}$ is the least symplectic eigenvalue of the partially-transposed $\textbf{V}^\prime$ of the $\textbf{V}$, associated with the selected bipartition, obtained by neglecting the rows and columns of the uninteresting mode,
\begin{equation}
\textbf{V}^\prime=\left( \begin{array}{cccc}
\mathcal{M}&\mathcal{N}\\
\mathcal{N}^T&\mathcal{M}^\prime
\end{array}\right), 
\end{equation}
and $\sigma=\mathrm{det}\mathcal{M}+\mathrm{det}\mathcal{M}^\prime-2\mathrm{det}\mathcal{N}$.

\begin{figure}[ht]
\centering
\subfloat[]{
\includegraphics[width=2.8in]{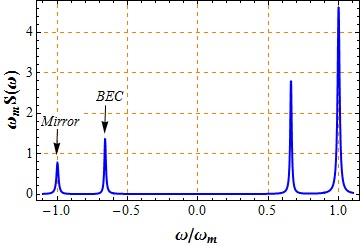}
\label{fig1a}}\\
\subfloat[]{
\includegraphics[width=2.8in]{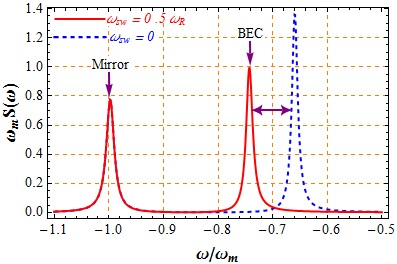}
\label{fig1b}}

\caption{(Color online) (a) Normalized cavity output power spectrum versus the normalized frequency $\omega/\omega_m$. (b) Stokes sideband of cavity output spectrum for two different values of collision parameters $\omega_{sw}=0$ (dashed line), and $\omega_{sw}=0.5 \omega_{R}$ (solid line) with the laser pump strength $50$mW, the cavity has a length $L=1$ mm, a wavelength of $\lambda=1080$nm with finesse $\mathcal{F}=3\times 10^4$, and the damping rate $\gamma_m=\pi c/L\mathcal{F}$, cavity detuning $\Delta=- \omega_{m}$. The end mirror of the cavity with mass $m = 50$ng oscillates with the natural frequency $\omega_{m}=2\pi \times 10^7$Hz at the temperature $T_{m}=0.04$K. The recoil frequency of BEC is $\omega_R/2 \pi=3.57\times10^3$ with dissipation of collective density excitations $\gamma_{c}=0.001\kappa$,  and temperature $T_{c}=1 \mu$K.} 
\label{fig1}
\end{figure}

\begin{figure}[ht]
\centering
\includegraphics[width=3in]{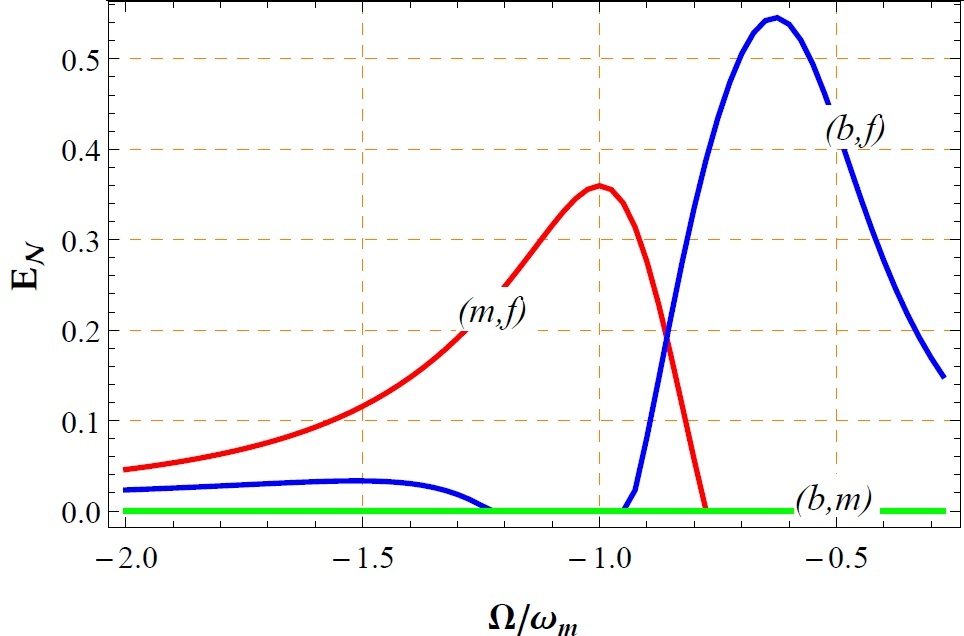}
\caption{(Color online) Logarithmic negativity $E_{\mathcal{N}}$ of the CV bipartite system formed by the moving mirror ($m$), BEC mode ($b$) and a single cavity output field ($f$) versus the central frequency of the detected output mode $\Omega/\omega_{m}$. The other parameters are the same as in Fig \ref{fig1}. The mechanical and BEC modes are significantly entangled only with the first Stokes sideband at $\Omega=-\omega_{m}$ and $\Omega=-\omega_B \simeq-0.6\omega_{m}$ respectively.}
\label{fig2}
\end{figure}

Firstly, we investigate the quantum state transfer from BEC and mechanical modes to output field in one cavity. Here beside the generated Stokes and anti-Stokes motional sidebands by means of mechanical resonator, the motion of collective modes of BEC can generate Stokes and anti-Stokes sidebands, consequently modifying the cavity output spectrum. Therefore it may be nontrivial to specify which is the optimal frequency bandwidth of the output cavity field that carries most of entanglement generated within the cavity. The output cavity field spectrum associated with the photon number fluctuations $S{(\omega)=\langle\delta{a^{out}(\omega)}^\dagger\delta{a^{out}(\omega)}\rangle}$ is shown in Fig.\ref{fig1a}, where we have considered a parameter regime with the laser pump power $50$mW, the cavity has a length $L=1$ mm, a wavelength of $\lambda=1080$ nm with finesse $\mathcal{F}=3\times 10^4$, and the damping rate $\pi c/L\mathcal{F}$, cavity detuning $\Delta=- \omega_{m}$. The end mirror of the cavity with mass $m = 50$ng oscillates with the frequency $\omega_{m}=2\pi \times 10^7$Hz at the temperature $T_{m}=0.04$K. The recoil frequency of BEC is $\omega_R/2 \pi=3.57\times10^3$ with dissipation of collective density excitations $\gamma_{c}=0.001\kappa$,  and temperature $T_{c}=1 \mu$K. Fig. \ref{fig1a} shows the power spectrum of output field against the normalized frequency has four peaks which are resonance with moving mirror and BEC Bogoliubov modes. In Fig. \ref{fig1b} we plot the spectrum of the output field for two different values of atomic collision $\omega_{sw}=0$, and $\omega_{sw}=0.5$. Fig. \ref{fig1b} shows the atomic interaction makes shift in the cavity resonance frequency and consequently reduces the cavity field intensity, and hence the decrease of the cavity output optical field would be a direct measure of the atom-atom interaction of the BEC.
In order to study the BEC and vibrating mirror and BEC Bogoliubov entanglement with output optical field, we consider logarithmic negativity $E_\mathcal{N}$ given by Eq. (\ref{EN}) for $\textbf{V}^{\prime}$.
In Fig. \ref{fig2} the BEC and mirror entanglement with output optical field is plotted versus $\Omega/\omega_m$. When the bandwidth is not too large ($\varepsilon=\omega_m \tau =10$), the mechanical and BEC modes are significantly entangled with only $\Omega=-\omega_{m}$ and $\Omega=-\omega_B \simeq-0.6\omega_{m}$ respectively. Also by comparing with previous works, the entanglement of the output mode is significantly larger than its intracavity counterpart \cite{dalafijpb,PhysRevA.78.032316}. So this analysis shows that it is possible to evaluate the entanglement properties of multipartite optomechanical system composed of BEC atoms and output modes for quantum communication application involves the manipulation of travelling optical fields.
In the next section, we use two of this system for generation entanglement between remote modes. Each system is initially composed of a pair of independent tripartite entangled states, one possesses by Alice and another by Bob. Now, Alice and Bob who have located at remote sites can utilize this system to prepare tripartite state, and each of them shares their optical mode to Charlie. The output optical field will be manipulated by Charlie to creating swapping entanglement between two remote subsystems.


\section{Entangling two remote non-interacting systems}\label{entanglement}
\subsection{Swapping entanglement protocol}\label{swapping ent}
In order to generate an entanglement state of two initially uncorrelated distant systems, we employ entanglement swapping \cite{doi:10.1142/S123016121350011X}, and apply it to the case when the two remote sites possess each a CV optomechanical system composed of BEC atoms(see Fig. \ref{f1}). In a such hybrid optomechanical system, an ensemble of ultracold atoms inside an optomechanical cavity can couple to the cavity field, where the excitation of the BECs plays the role of the mechanical mode of the mirror in an optomechanical system. Like standard optomechanical systems, the coherent motion of ultracold atoms causes a nonlinearity, and the atomic collisions change the resonance frequency of the cavity\cite{PhysRevA.87.013417}. In our systems, atomic and mechanical coupling with optical field are the basic ingredient for generating entanglement between the two distant systems. The whole system is composed of two seperable and independent
tripartite bosonic modes, where each of them possessed by Alice and Bob. Alice and Bob are located at remote sites and prepare a tripartite state, and each shares one modes (optical modes) with Charlie, who is located for simplicity halfway between them. Charlie can then perform a Bell measurement on the two optical modes and consequently entangle the two distant BEC, and mechanical modes of moving mirrors by means of CV entanglement swapping. 

Here, we consider an ideal Bell measurement protocol applied to the last output optical modes of each cavity with balanced transmissivity $T=1/2$, which allows Charlie to entangle the remote modes. We assume Alice and Bob initially possess the same CV tripartite state and each of them shares one travelling output optical
mode with Charlie, who is located for simplicity halfway between them. The covariance matrix of the tripartite bosonic modes at each site is fully characterized in Eq. (\ref{vmat}). The $12\times 12$ covariance matrix of the whole system, composed of two independent tripartite bosonic modes, can be written in the blockform
\begin{equation}
\textbf{V}=\left( \begin{array}{cccc}
\textbf{A}&\textbf{C}\\
\textbf{C}^T&\textbf{B}^{(2)}
\end{array}\right), 
\end{equation}
where the $8\times 8$ matrix $\textbf{A}$ is the reduced CM of the BECs, and mechanical modes, and 
\begin{equation}
\textbf{C}=\left( \begin{array}{cc}
\textbf{C}_1&\textbf{C}_2
\end{array}\right), 
\end{equation}
is a rectangular $(8\times 4)$ real matrix, describing the correlations between the first two remote modes and the output optical modes, and
\begin{equation}
\textbf{B}^{(2)}=\left( \begin{array}{cccc}
\textbf{B}_1&\textbf{D}\\
\textbf{D}^T&\textbf{B}_2
\end{array}\right), 
\end{equation}
is the reduced CM of the output optical modes received by Charlie (labelled by 1 and 2). For more simplification, we write down the reduced covariance matrix $\textbf{B}^{(2)}$ of output optical modes by setting 
\begin{eqnarray}
\mathbf{B}_1&&:=\left( \begin{array}{cccc}
\alpha_1 & \alpha_3\\
\alpha_3 & \alpha_2\\
\end{array}\right), \,\,
\mathbf{B}_2:=\left( \begin{array}{cccc}
\alpha_1' & \alpha_3'\\
\alpha_3' & \alpha_2'\\
\end{array}\right),\nonumber\\
\mathbf{D}&&:=\left( \begin{array}{cccc}
\beta_1 & \beta_3\\
\beta_3 & \beta_2\\
\end{array}\right). \nonumber
\end{eqnarray}
Now, the output optical modes to Charlie are passed to a beam spliter of transmissivity $T$. Then, Charlie applies a Bell-like measurement on the received optical modes by appliying homodyne detections on them as depicted in Fig. \ref{fig3}. The state of remaining remote quadrature fluctuations of BECs and mechanical modes of two remote systems would be a Gaussian state with CM of the form
\begin{equation}\label{CMT}
\bm{\mathcal{V}}=\textbf{A}-\frac{1}{\mathrm{det}\mathbf{\Gamma}}\sum_{i,j=1}^2 \mathbf{C}_i \mathbf{K}_{ij}\mathbf{C}_j^T:=\left( \begin{array}{cccc}
\bm{\mathcal{A}} & \bm{\mathcal{C}} \\
{\bm{\mathcal{C}}}^T&\bm{\mathcal{B}}
\end{array}\right), 
\end{equation}
where
\begin{eqnarray}
\mathbf{K}_{11}&&:=\left( \begin{array}{cccc}
(1-T)\gamma_2 & \sqrt{T(1-T)}\gamma_3\\
\sqrt{T(1-T)}\gamma_3 &T\gamma_1\\
\end{array}\right),\nonumber\\
\mathbf{K}_{22}&&:=\left( \begin{array}{cccc}
T\gamma_2 & -\sqrt{T(1-T)}\gamma_3\\
-\sqrt{T(1-T)}\gamma_3 &(1-T)\gamma_1\\
\end{array}\right),\nonumber\\ 
\mathbf{K}_{12}&&=\mathbf{K}_{21}^T:=\left( \begin{array}{cccc}
 -\sqrt{T(1-T)}\gamma_2 &(1-T)\gamma_3\\
-T\gamma_3 & -\sqrt{T(1-T)}\gamma_1\\
\end{array}\right),\nonumber
\end{eqnarray}
with 
\begin{eqnarray}
\mathbf{\Gamma}:=\left( \begin{array}{cccc}
\gamma_1 & \gamma_3\\
\gamma_3 & \gamma_2\\
\end{array}\right),\nonumber
\end{eqnarray}
and 
\begin{eqnarray}
\gamma_1  &:=& (1-T)\alpha_1+T \alpha_1'-2\sqrt{T(1-T)}\beta_1,\nonumber\\
\gamma_2  &:=&T\alpha_2+(1-T) \alpha_2'+2\sqrt{T(1-T)}\beta_2,\nonumber\\
\gamma_3  &:=&\sqrt{T(1-T)}(\alpha_3'-\alpha_3)-(1-T) \beta_3+T\beta_4.\nonumber
\end{eqnarray}
$\bm{\mathcal{A}}$ , $\bm{\mathcal{B}}$ describe the remaining BECs and mechanical modes after Bell-like detection, and $\bm{\mathcal{C}}$
referred to the cross-correlation elements. 
In order to study the bipartite entanglement between each two remote modes, we consider logarithmic negativity eq.(\ref{EN}),
where $\eta_{-}$ is the least partially-transposed symplectic eigenvalue of $\bm{\mathcal{V}}$, and $\sigma=\mathrm{det}\bm{\mathcal{A}}+\bm{\mathcal{B}}+2\bm{\mathcal{C}}$.
\subsection{Entanglement of the BECs and mechanical resonators by swapping}
\begin{figure}[ht]
\centering
\includegraphics[width=4.3 in]{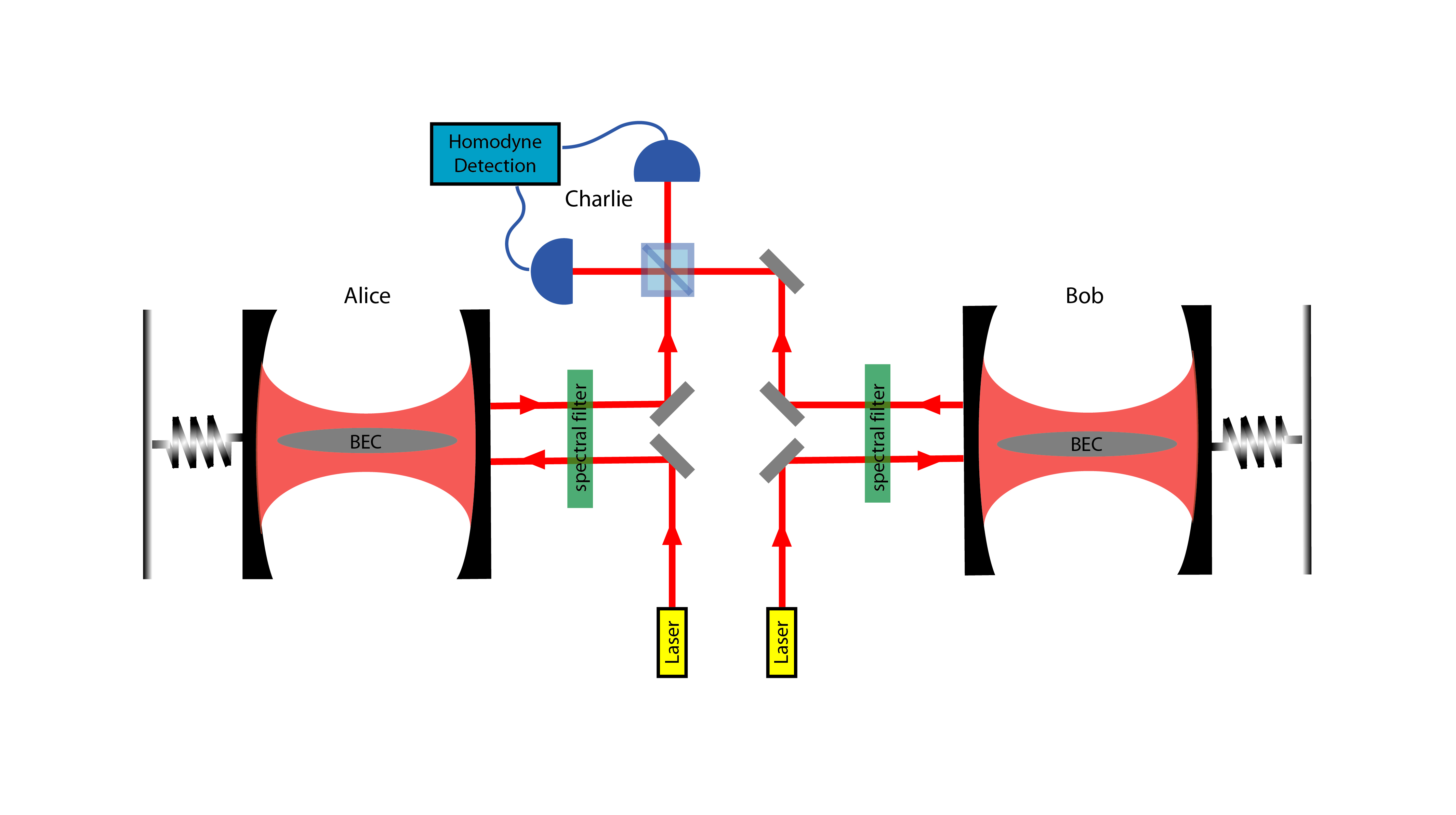}
\caption{(Color online) Scheme of the entanglement swapping protocol for entangling of two distant hybrid optomechanical cavity containing BEC atoms by applying a Bell-like detection.}
\label{fig3}
\end{figure}

\begin{figure*}
\centering
\includegraphics[width=3.2in]{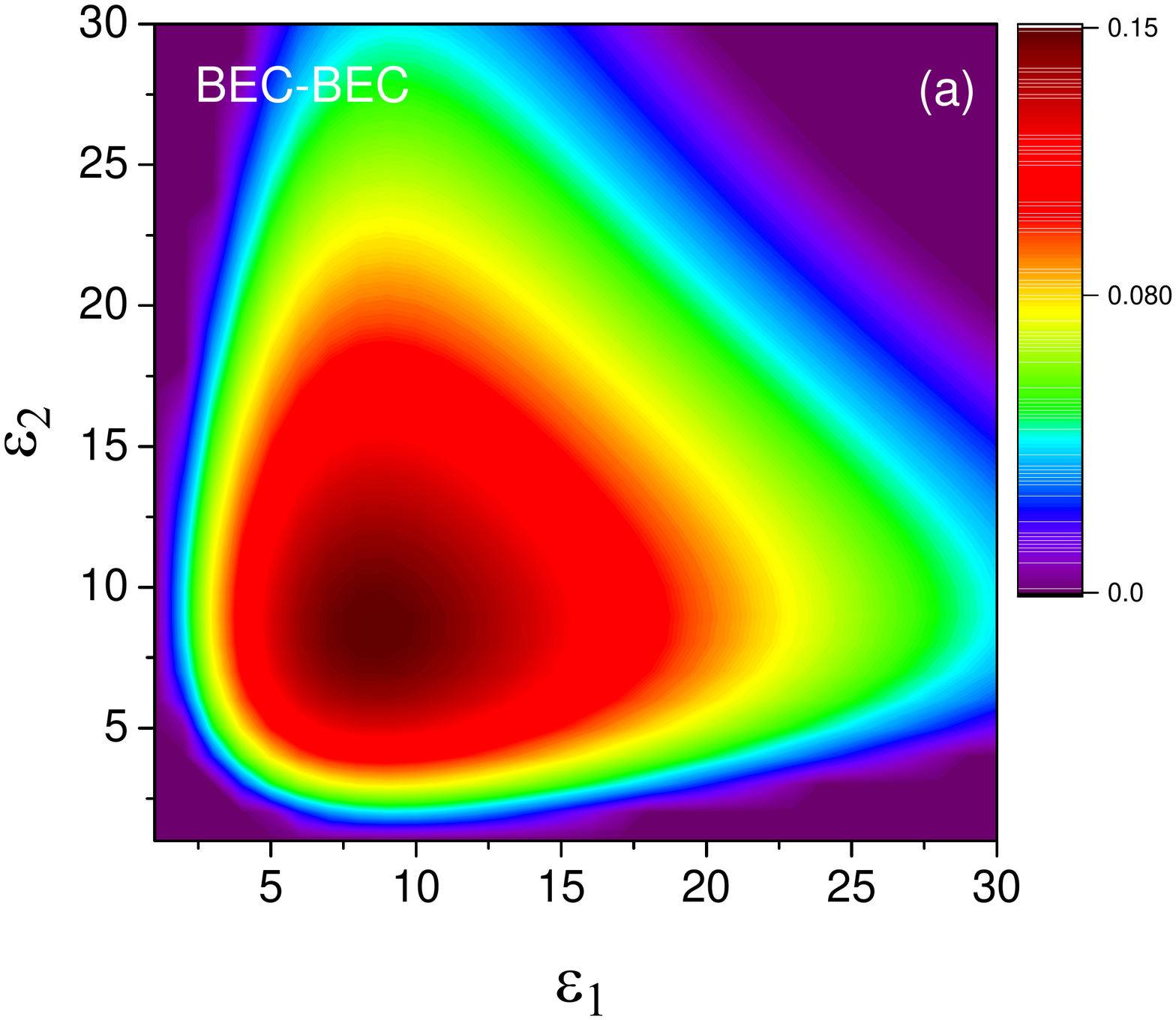}
\includegraphics[width=3.2in]{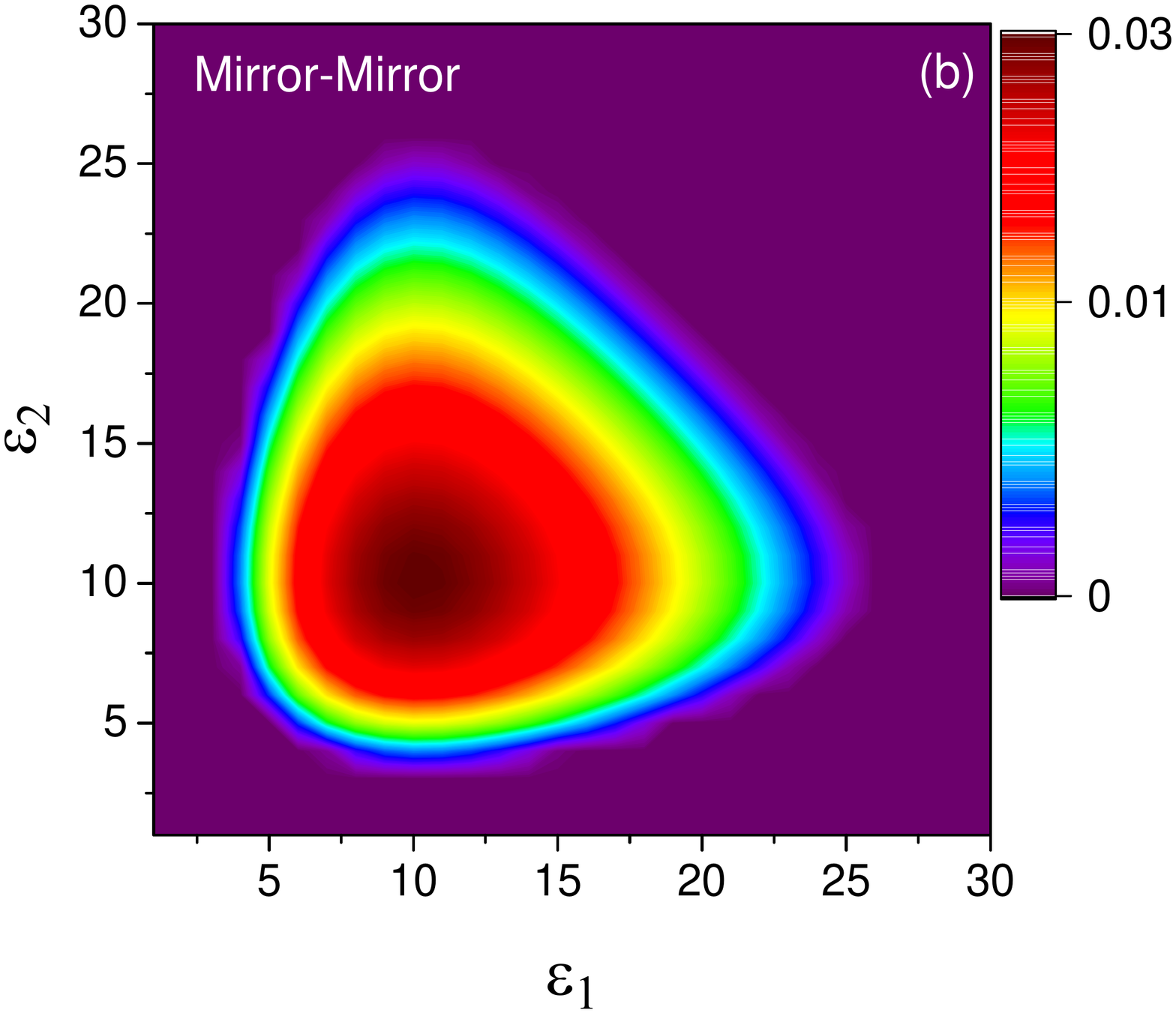}
\includegraphics[width=3.2in]{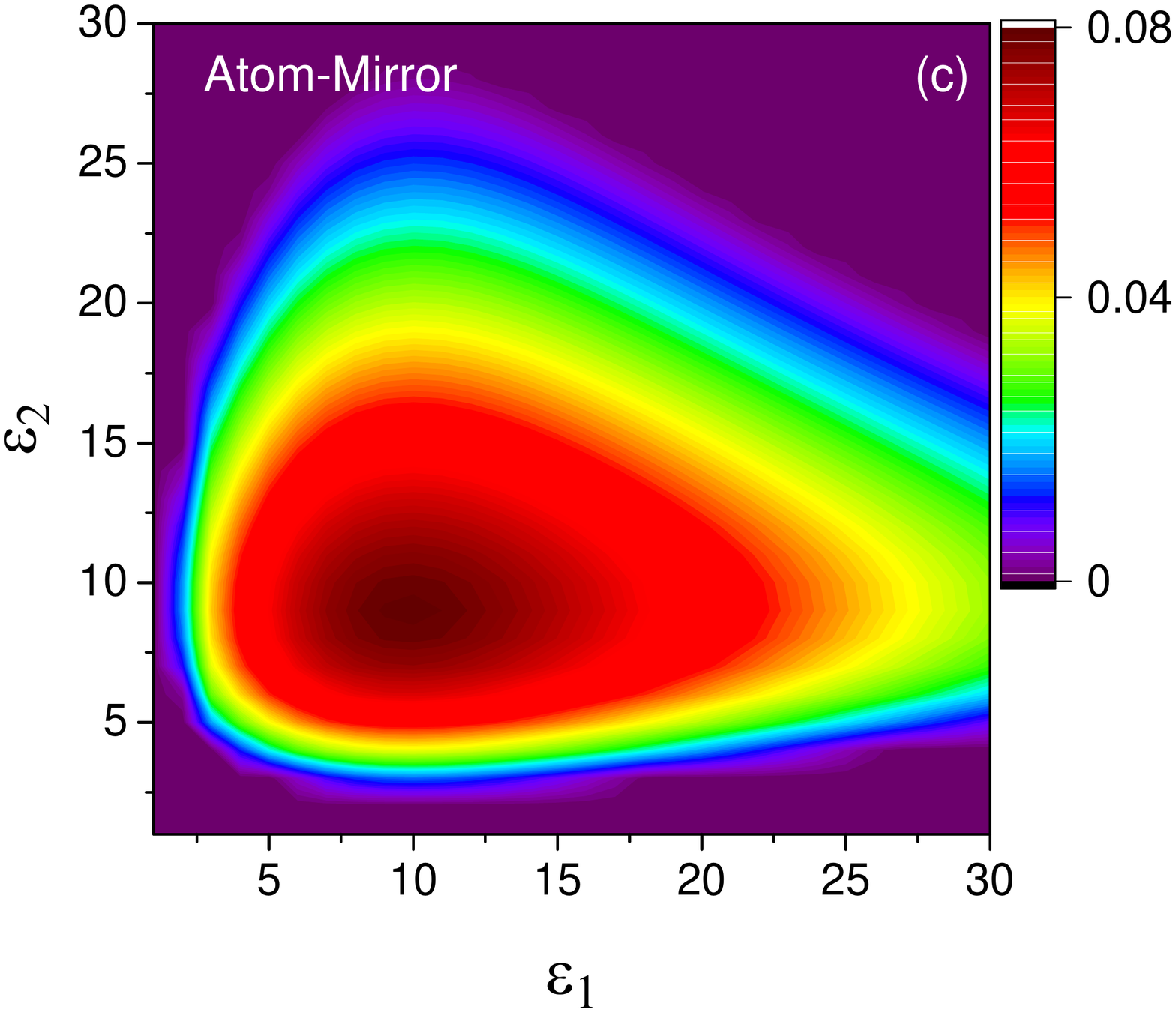}
\caption{(Color online) Logarithmic negativity $E_\mathcal{N}$ of the (a) ($b_A, b_B$) , (b) ($m_A, m_B$) , and (c) ($m_A, b_B$) modes versus inverse bandwidth $\varepsilon$.  The other parameters are the same as in Fig. \ref{fig1}.} 
\label{fig7}
\end{figure*}

\begin{figure*}
\centering
\subfloat[]{
\includegraphics[width=0.35\textwidth]{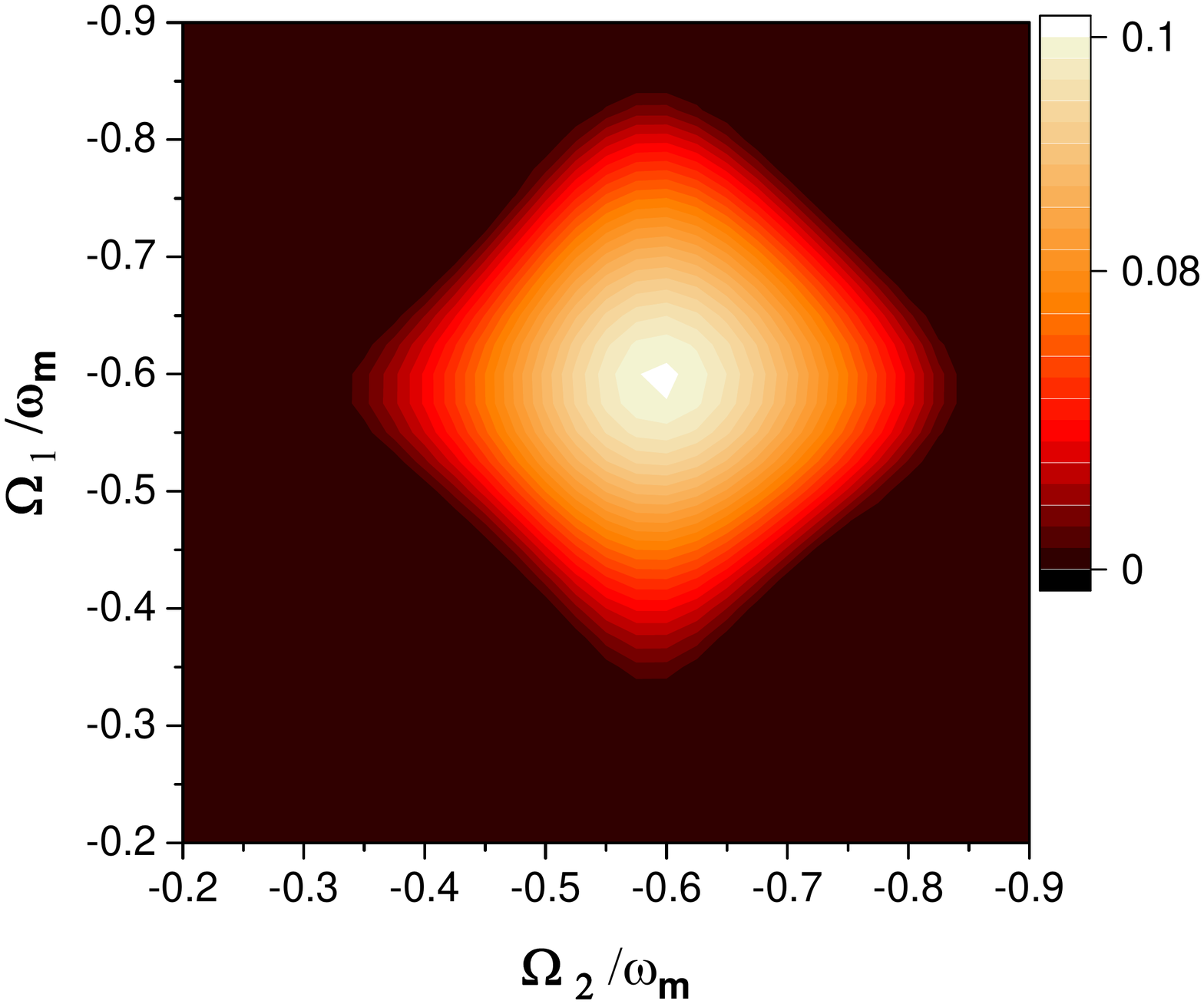}
\label{fig4}}
\subfloat[]{
\includegraphics[width=0.35\textwidth]{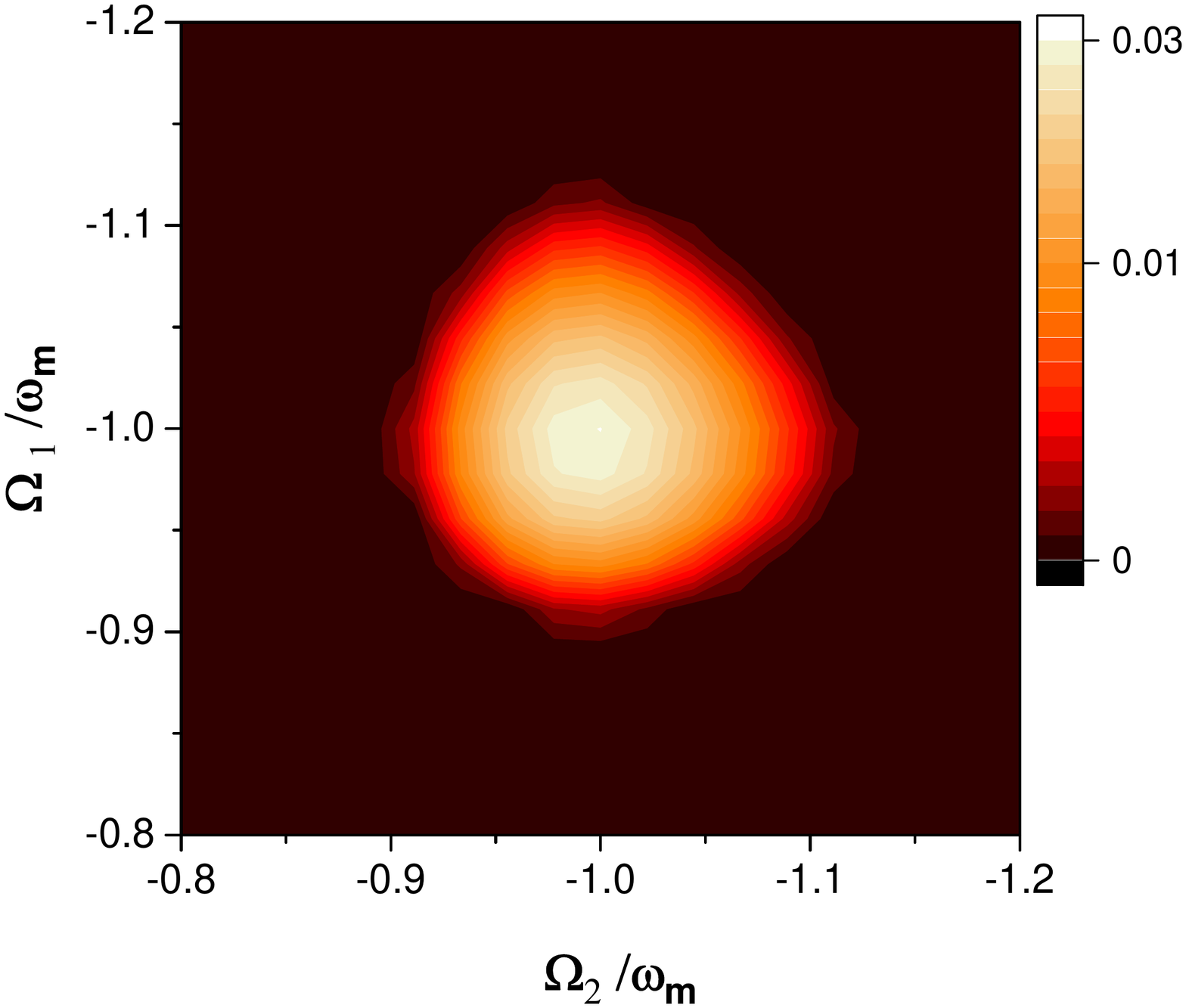}
\label{fig5}}
\subfloat[]{
\includegraphics[width=2.5in,height=2in]{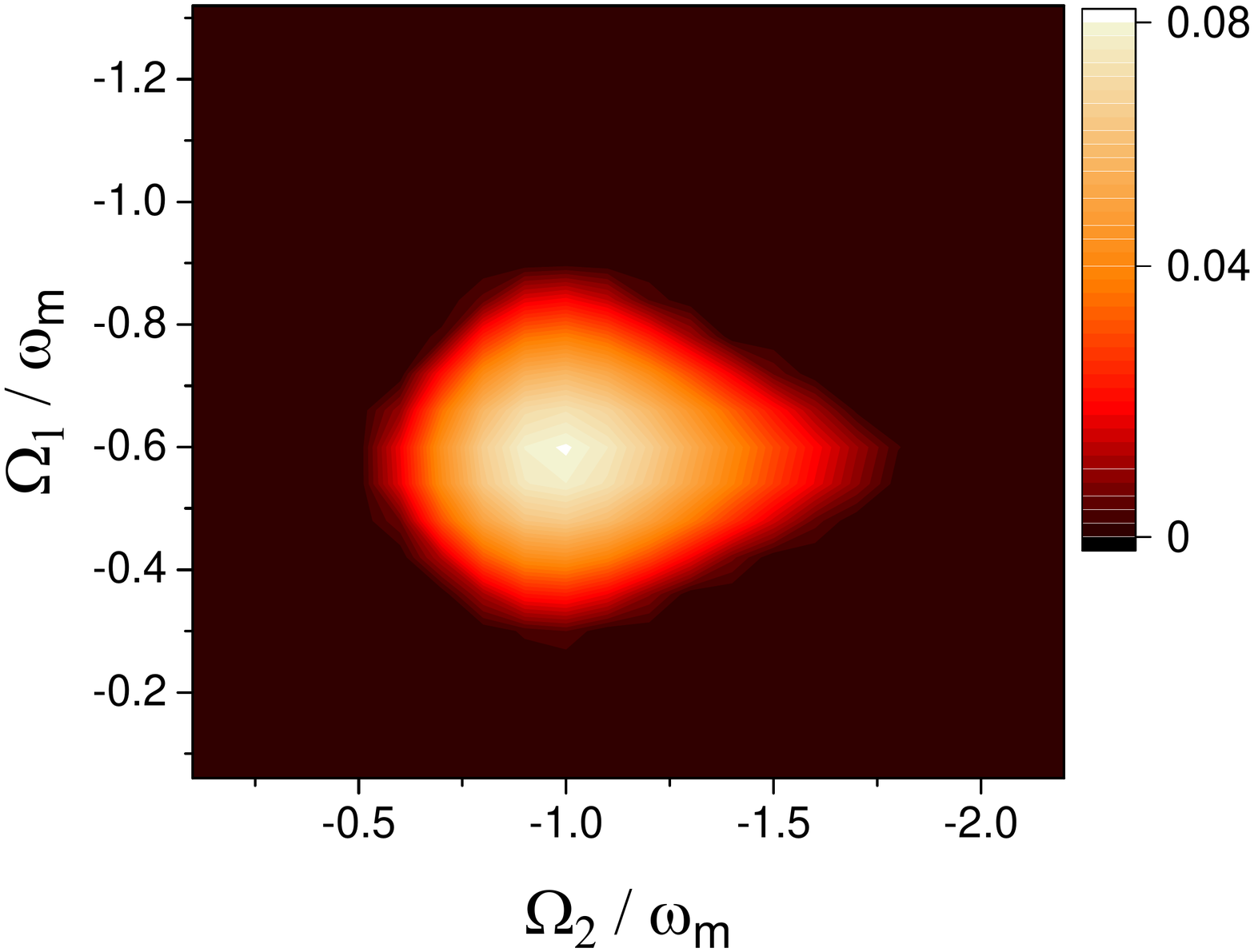}
\label{fig6}}
\qquad
\subfloat[]{
\includegraphics[width=0.35\textwidth]{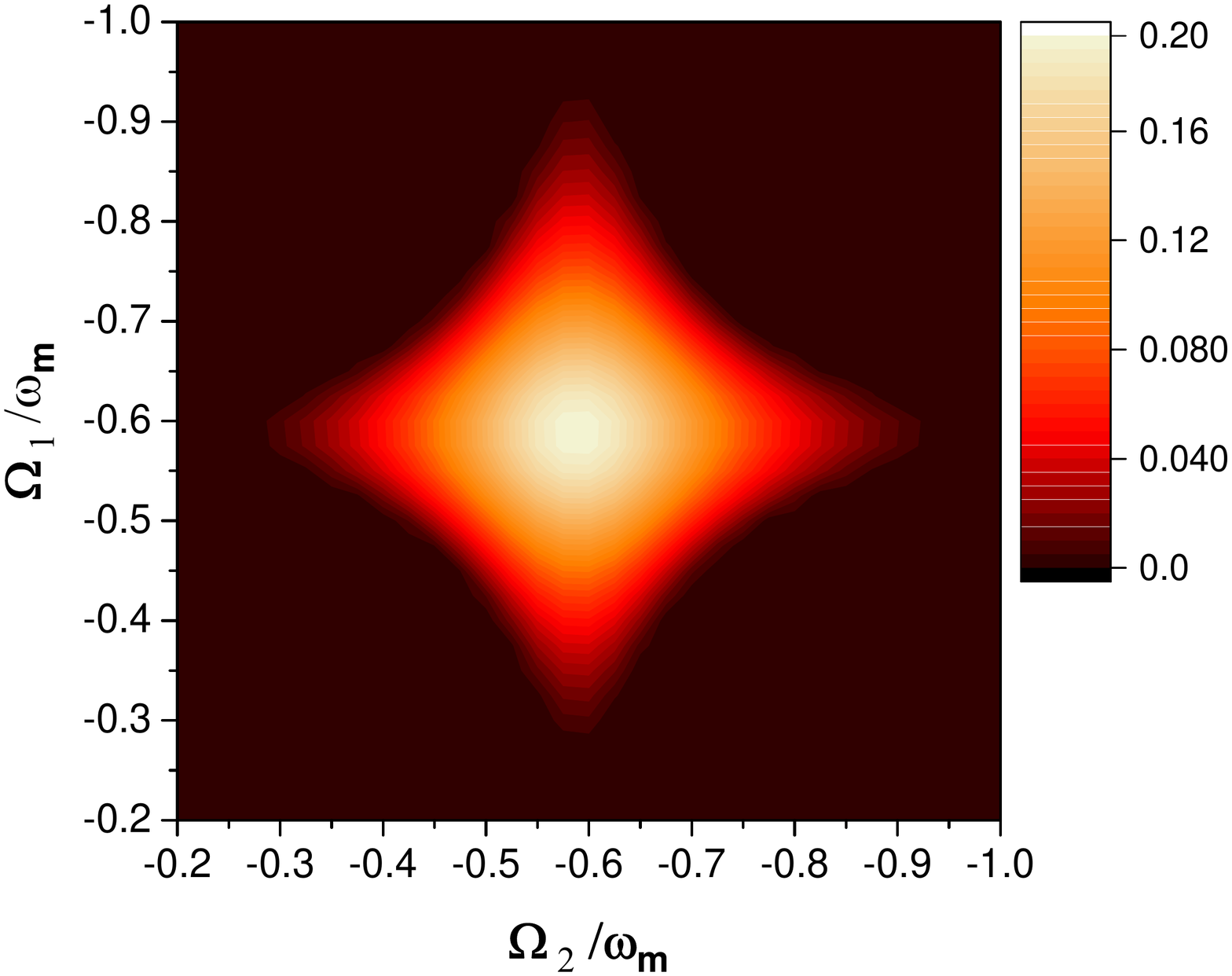}
\label{fig4g}}
\subfloat[]{
\includegraphics[width=0.35\textwidth]{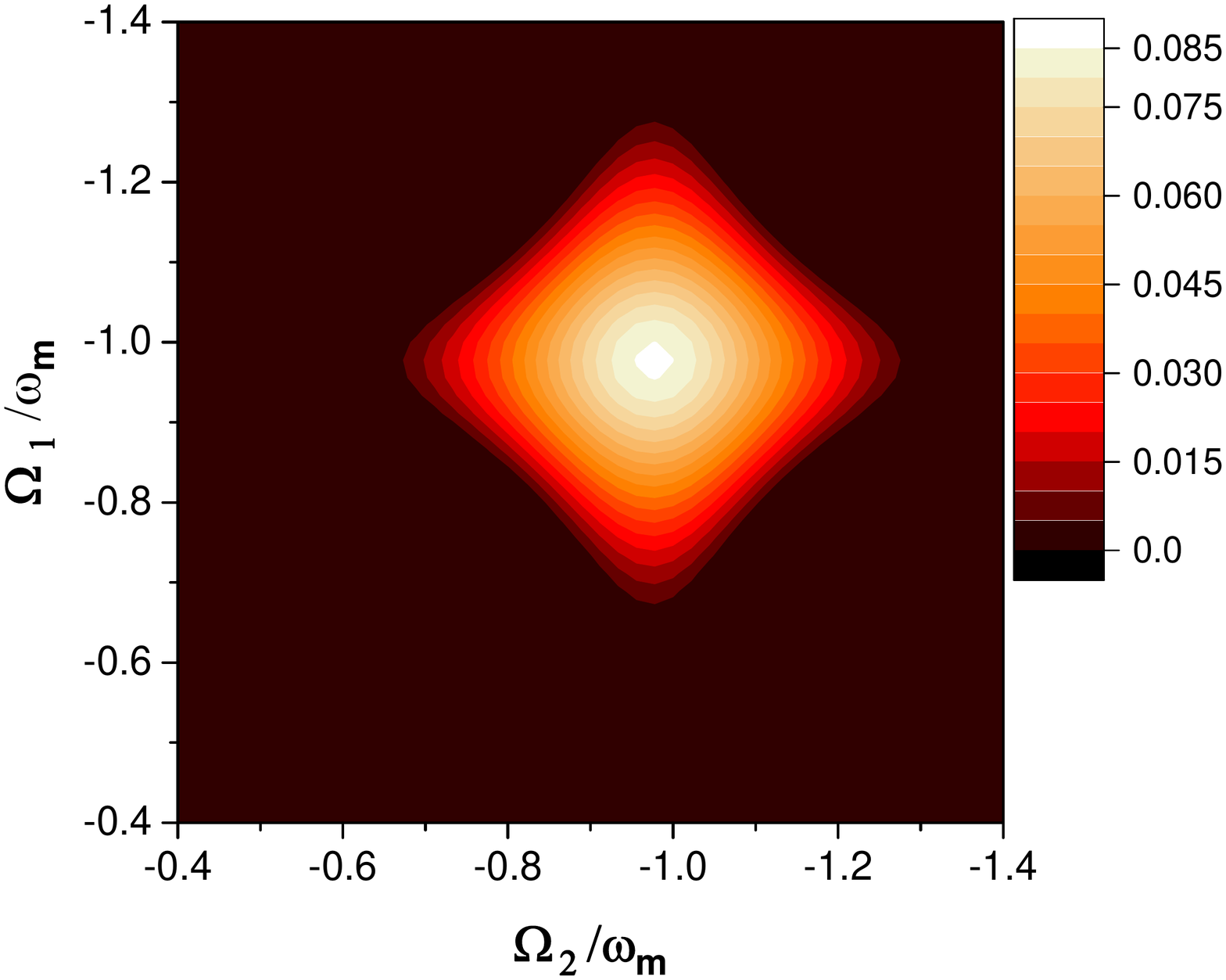}
\label{fig5g}}
\subfloat[]{
\includegraphics[width=0.35\textwidth]{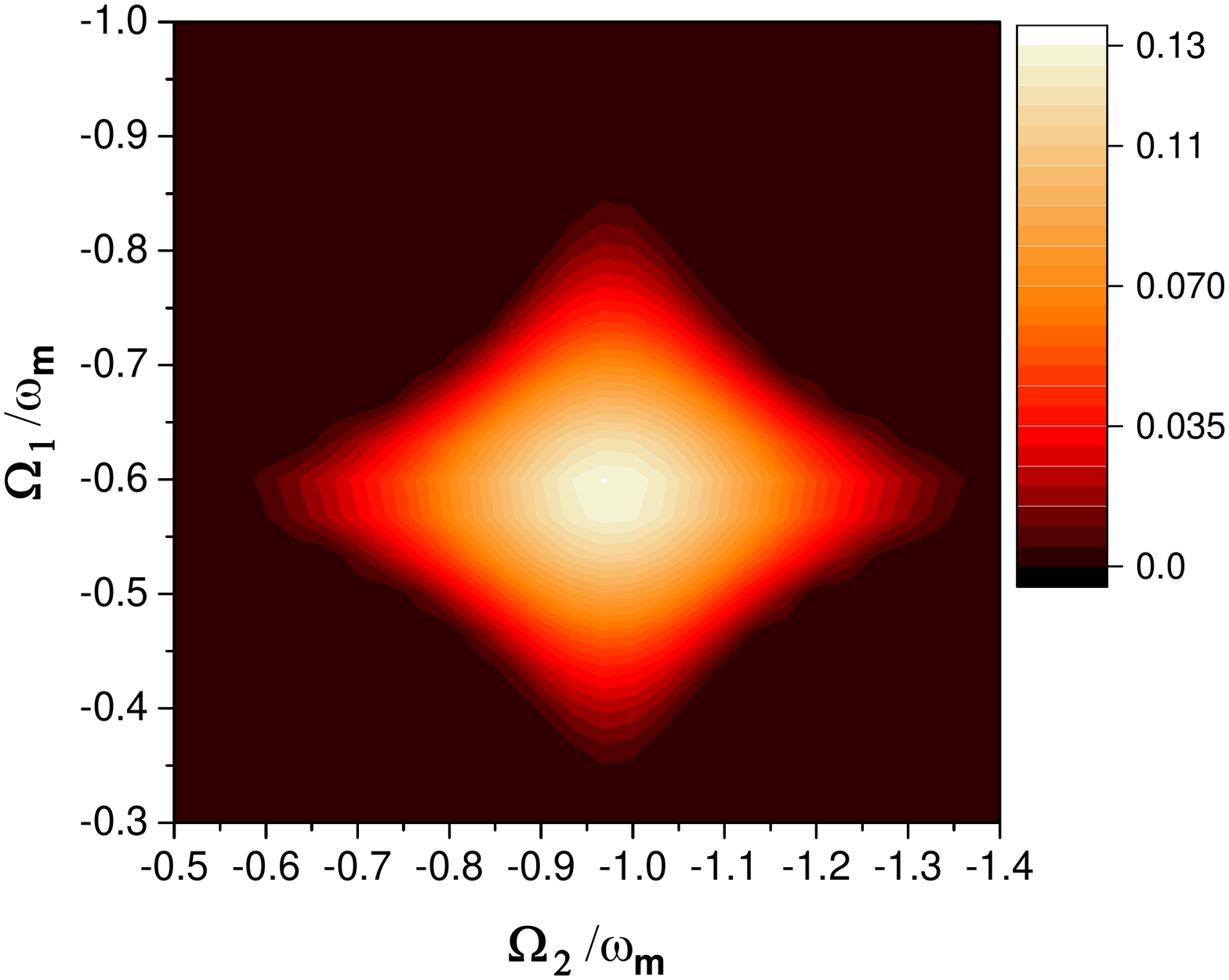}
\label{fig6g}}   
\caption{(Color online) Logarithmic negativity $E_\mathcal{N}$ of the (a) ($b_A, b_B$), (b) ($m_A, m_B$), and (c) ($m_A, b_B$) modes versus the central frequency of the detected output mode $\Omega_i/\omega_m$ (i=1,2). The maximum of entanglement occurs around the stokes $\Omega_i=-\omega_B\simeq-0.6\omega_m$ for BECs, $\Omega_i=-\omega_m$ for mirrors. (d-f) same to (a-c), but when the coupling constants have been turned off: (d) (${g_A}={g_B}=0$), (e) ($G_A=G_B=0$), and (f) ($G_A={g_B}=0$).The other parameters are the same as in Fig. \ref{fig2}.}  
\label{fig456}
\end{figure*}

\begin{figure*}
\centering
\subfloat[]{
\includegraphics[width=0.34\textwidth]{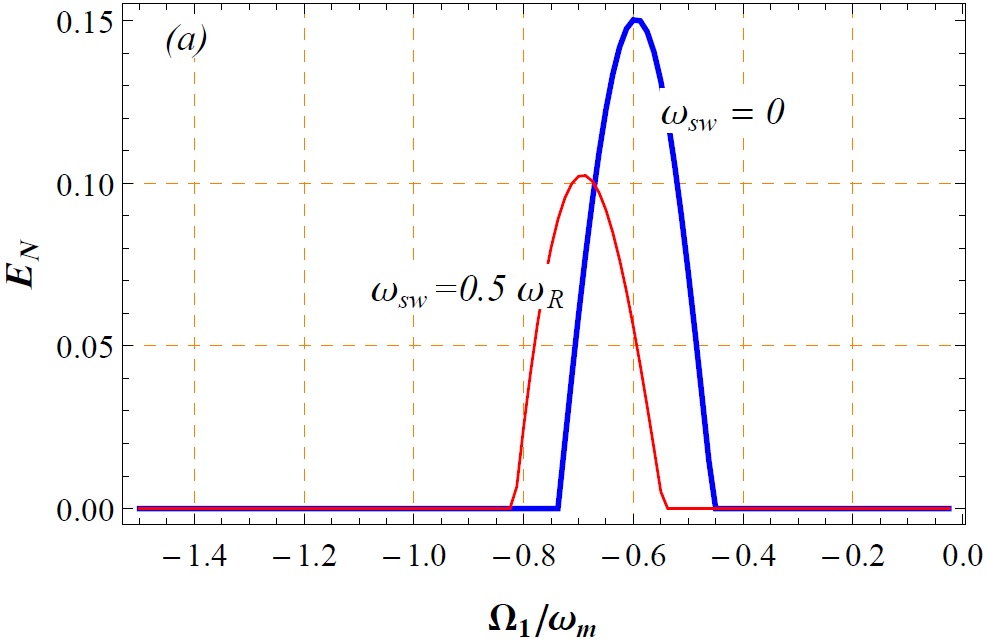}
\label{fig9a}}
 \subfloat[]{
\includegraphics[width=0.34\textwidth]{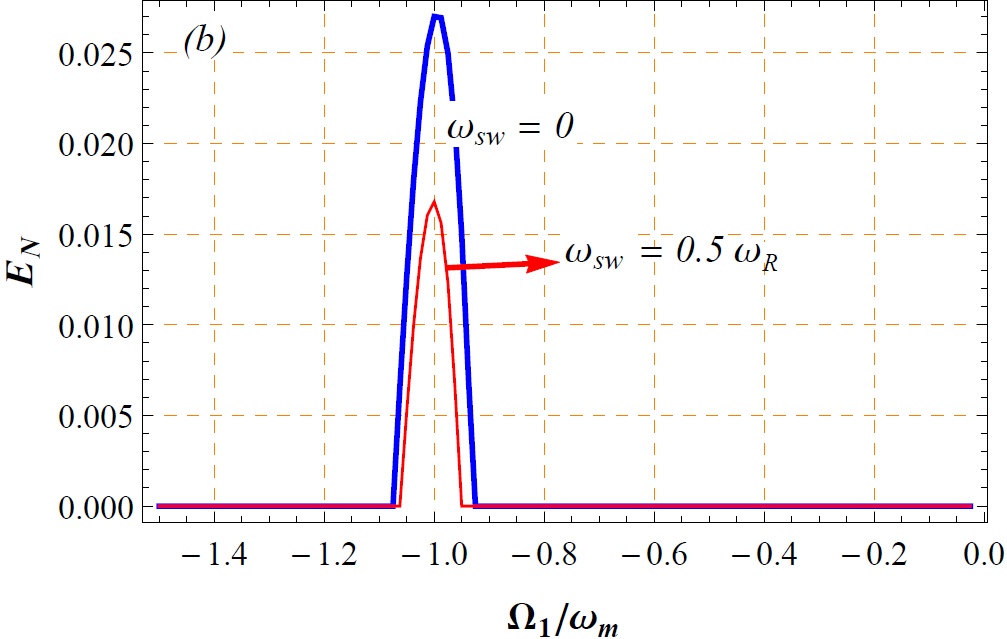}
\label{fig9b}}
 \subfloat[]{
\includegraphics[width=0.34\textwidth]{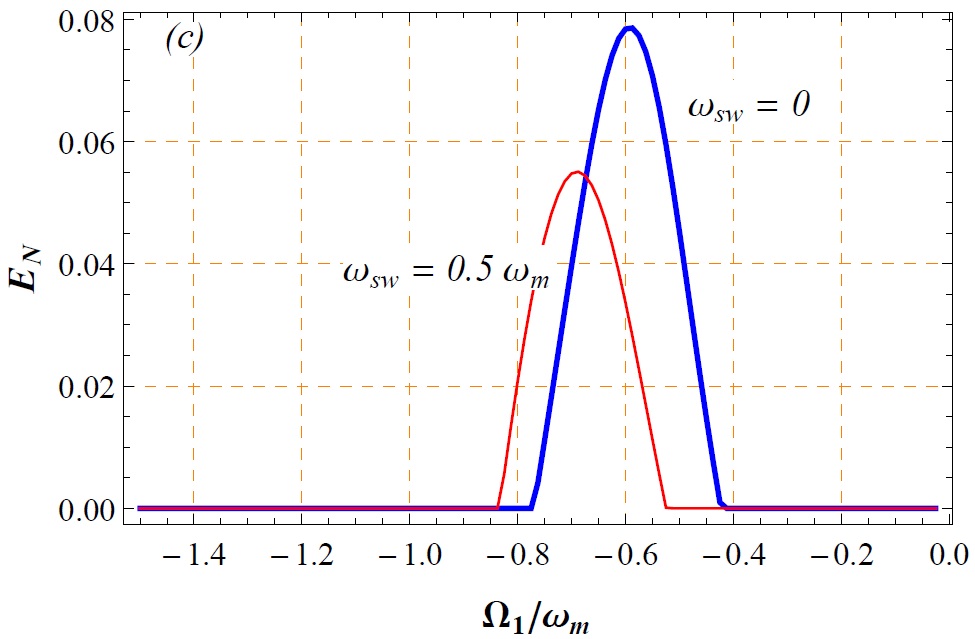}
\label{fig9c}}
\caption{(Colour on line) Entanglement between (a) ($b_A, b_B$), (b) ($m_A, m_B$), and ($b_A, m_B$)  modes against the central frequency of the detected output mode $\Omega_1/\omega_m$ for two different values of collision parameters $\omega_{sw}=0$, and $\omega_{sw}=0.5 \omega_R$ of BEC atoms.}
\label{fig9}
\end{figure*}

Here we now utilize the results obtained in the section \ref{swapping ent} to show that how we can entangled of two distant remote modes in two independent cavity which are physically separated by implementation of Bell measurement. The scheme is shown in Fig \ref{fig3}. The parameters of system must be chosen in such a way that the tripartite system formed by BECs ($b_A$, $b_B$), mechanical modes ($m_A$, $m_B$) and optical fileds ($f_A$, $f_B$) satisfy the Routh-Hurwitzr stability criterion \cite{gradshteyn1980tables}. Therefore, we have to determine an experimentally achievable parameter set in which such conditions are satisfied so that the proposed generalized swapping protocol can be successfully implemented. The entanglement between BEC and output field and also mechanical resonator and output optical field employed for the Bell measurement. By appropriately choosing the detuning  and filtering the appropriate output modes, we can satisfy stationary condition and generate entanglement between non-interacting systems. We generalized the model to the symmetric case of initially identical state at Alice and Bob. One can modifies the coupling constant of each subsystem and changes the model to the same or different subsystems.  
Firstly, we sketched Fig. \ref{fig7} for finding the optimal filtering bandwidth. Fig. \ref{fig7} shows the logarithmic negativity $E_\mathcal{N}$ between different bipartite modes as a function of the normalized filtering bandwidth $\varepsilon=\omega_m \tau $, obtained after the Bell measurement. It can be seen from the figure that logarithmic negativity achieves its maximum value at the optimal value $\varepsilon\simeq10$ for ($b_A$, $b_B$), ($m_A$, $m_B$), and ($f_A$, $f_B$), but It was not seen any entanglement between the subsystems - ($m_A, b_A$), ($m_B, b_B$) - at each node. This means that, in practice, by appropriately filtering the output field one realize an effective entanglement generation because of Bell measurement at different nodes.
A more interesting situation is depicted in Fig. \ref{fig456} which shows how entanglement between the remote systems depends on the frequency of detected output modes.
In Fig. \ref{fig4} the entanglement between the Bogoliubov modes of BECs ($b_A, b_B$) inside the two cavity has been plotted versus the frequency of the detected output modes $\Omega_i/\omega_m$ (i=1,2). This figure shows that the maximum entanglement between the Bogoliubov modes transfer around the $\Omega_1=\Omega_2=-\omega_B$ which is correspond to $\Omega_1=\Omega_2\simeq-0.6\omega_m$. So according to the Fig. \ref{fig2} the maximum entanglement between the output field and BEC causes the maximum entanglement between ($b_A, b_B$) modes after the Bell detection for filtering inverse bandwidth $\varepsilon\simeq10$. In Fig. \ref{fig5} the entanglement between the mirror modes ($m_A,m_B$) of the two cavity has been plotted versus the central frequency of the detected output mode $\Omega_i/\omega_m$ (i=1,2). The maximum of entanglement occurs around the $\Omega_1=\Omega_2=-\omega_m$. In fact, Figs. \ref{fig4}, \ref{fig5} show the optomechanical entanglement is large when we drive the cavity Bell mode (output filed) with a blue-detuned laser ($\Omega_i=-\omega_m$ for mechanical resonator and $\Omega_i=-\omega_B$ for BEC modes (i=1,2)). Also in Fig. \ref{fig6} the entanglement between the mirror and BEC modes of the two cavity ($m_A, b_B$) has been plotted versus the central frequency of the detected output mode $\Omega_i/\omega_m$ (i=1,2). The maximum of entanglement occurs around the $\Omega_1=-\omega_B\simeq-0.6\omega_m$ and $\Omega_2=-\omega_m$ for mirror and BEC respectively. Figs. (\ref{fig456}\subref{fig4g}-\subref{fig6g}) are the same the Figs.  (\ref{fig456}\subref{fig4}-\subref{fig6}), when the coupling constant of other subsystems is zero. It qualitatively has shown that the manner of entanglement is very similar to Figs. (\ref{fig456}\subref{fig4}-\subref{fig6}), and generally we can study the generation of entanglement between same or different remote modes by turning off (on) the coupling constant between subsystems. For instance, by turning off the coupling constant of mirror in the Alice system ${g_A}=0$, and BEC in the Bob system $G_B=0$, we can choose different remote systems for some applications of quantum informations like quantum teleportation. 
Fig.\ref{fig9} we plot the entanglement between (a) ($b_A, b_B$), (b) ($m_A, m_B$), and ($b_A, m_B$) modes against the central frequency of the detected output mode $\Omega_1/\omega_m$ for two different atomic collisions $\omega_{sw}$. This figure shows s-wave scattering makes the shift of resonance frequency and causes the the entanglement between ($b_A, b_B$), and ($b_A, m_B$) to decrease. Hence the decrease of the cavity output intensity causes the weaker entanglement between BECs after the Bell measurement, and also between two moving mirrors. But against the Fig.\ref{fig9} (a) and (c), in Fig.\ref{fig9}(b) the atom-atom interaction dose not change the cavity resonance frequency.
\section{conclusions}\label{conclusion}
In this paper firstly we have investigated the entanglement between the different bipartite systems formed by BEC, mechanical, and output optical modes, which is important from a practical point of view because any quantum-communication application involves the manipulation of travelling optical fields. It has been shown that the Stokes output mode is strongly entangled with the BEC, and mechanical modes, and also there is no entanglement between the mechanical and BEC modes. Secondly we have shown that entanglement swapping protocol can be implemented using two distant hybrid optomechanical system. We have considered two remote optomechanical cavity composed of one dimensional ultracold atoms inside the cavity and performed Bell measurement for entangling two same or different remote systems. It was shown that Bell measurement entangled all of the subsystems except the BEC and mechanical mode in the position of each node. Also, we investigated the effect of atomic collision on the entanglement of different remote modes. It has been shown that an increase in the s-wave scattering frequency caused the entanglement of the remote systems to decrease. Besides, an increase in the s-wave scattering frequency caused shift the resonance frequency of cavity in the cases of $(b_A, b_B)$, and $(b_A, m_B)$.
\section*{Acknowledgements}
The authors wish to thank the office of graduate studies of the University of Kashan, and University of Hormozgan for their support. This work was supported by Council of University of Kashan by Grant Agreement No. 572801/4
\bibliography{BEC}
\bibliographystyle{apsrev4-1} 
\end{document}